\documentclass[preprint,12pt]{elsarticle}

\usepackage{amssymb}

\usepackage{amsthm}

\usepackage{tabularx} 
\usepackage{threeparttable} 
\usepackage[%
breaklinks=true,
colorlinks=true,
linkcolor=blue,
urlcolor=blue,
citecolor=blue
]{hyperref}

\usepackage{bm}
\usepackage{amsmath}
\usepackage{enumitem}

\journal{Some journal} 

\begin{document}

\begin{frontmatter}



\title{Application of Machine Learning for the Identification of 2D Colloidal Assemblies: A Case Study on Particles of Distinct Shapes}


\author[label1]{L. T. Khusainova}
\author[label2]{S. A. Kolegova}
\author[label1]{K. S. Kolegov}

\affiliation[label1]{organization={Astrakhan Tatishchev State University},
            addressline={20a Tatischev Str.},
            city={Astrakhan},
            postcode={414056},
            country={Russia}}

\affiliation[label2]{organization={Caspian Institute of Maritime and River Transport named after Admiral General F.M. Apraksin – the affiliation of “Volga State University of Water Transport”},
            addressline={17 Uritskogo Str. / 6 Nikolskaya Str. / 14 Fioletova Str.},
            city={Astrakhan},
            postcode={414000},
            country={Russia}}

\begin{abstract}
This work addresses the problem of identifying colloidal monolayer assemblies using particles of various shapes (two-dimensional coatings): spheres, ellipsoids, cuboids, and rods. The following classification of assemblies is considered: isolated particles, dimers, chains, clusters, and loops. The YOLO model was chosen as the identification method. Synthetic datasets were prepared for each of the four particle shapes to train the models. The paper discusses the application of models trained on synthetic data to experimental images. An analysis was carried out on the feasibility of using such models for recognizing configurations in real images. While recognition on artificial images is nearly perfect, tests on experimental images showed a significant deviation. The average error across all particle types was 43.1\%, but a considerable spread in values is observed: from 20\% for spheres to 58.5\% for cuboids, indicating the algorithm's selective sensitivity to object geometry. The created datasets and trained models are freely available for use. The corresponding modules have been integrated into the previously developed information system (\href{https://isanm.space/}{https://isanm.space/}). To further improve prediction results, it is necessary to prepare datasets based on experimental images.
\end{abstract}




\end{frontmatter}


\section*{Introduction}
\label{sec:Introduction}
The study of the morphology of colloidal structures is of not only theoretical but also applied interest. The self-assembly of densely packed colloidal particles, for example, into two-dimensional (2D) ordered arrays, is of key importance for nanotechnology. Such structures are used in the fabrication of photonic crystals, plasmonic devices, and biomimetic surfaces with specified properties (wetting, reflection, etc.) \cite{Lotito2017}. Theoretical studies make it possible to identify morphologies that, once realized in practice in artificially created materials and coatings, can yield their unique properties~\cite{Klatt2019}. The analysis of deposits in dried sessile droplets of biofluids is promising for medical applications related to disease diagnosis~\cite{Pal2024,Pal2025}.

Despite the large number of experimental techniques used to study and analyze colloidal systems, their application faces many key limitations. These include the low speed of manual image processing, the subjectivity of identification criteria, and so on. For example, the ambiguity in determining particle neighbors is one of the problems in the analysis of colloidal structures~\cite{Mickel2013}. When dealing not with spherical particles but with particles of arbitrary shape or their mixtures, defining a neighbor criterion becomes even more difficult. Machine learning, particularly modern computer vision architectures, opens up new opportunities in materials science~\cite{Sukhoverkhova2025}, including in the automated analysis of colloidal structures~\cite{Long2014,Carstensen2018,Newby2018,Boattini2019}. Unlike traditional methods that require manual processing and expert interpretation, it offers high speed and scalability, as well as high accuracy in morphology analysis. 

Subramanian et al.~\cite{Subramanian2021} developed an automated pipeline for mining and analysing microscopy images from the gold nanoparticle literature. They employed a Mask R-CNN model to segment nanoparticles and simultaneously classify their morphologies (sphere, rod, cube, and triangular prism) in a single forward pass, thereby avoiding error accumulation between segmentation and classification steps. Their work yielded a publicly available dataset of 4361 SEM/TEM images with automatically extracted size and morphology information. However, their approach focuses on the identification of individual particle shapes, rather than on the structural assemblies that these particles form. 
Here we consider one of the many possible classifications used in the analysis of particle assemblies. According to this classification, particle assemblies in a monolayer can be divided into the following types: single (isolated) particles, dimers, chains, loops, and clusters~\cite{Lotito2017,Lotito2019}. Such types of assemblies may be of interest, for example, in problems related to the fabrication of 1D and 2D photonic crystals~\cite{Shi2024}.

The above classification of colloidal assemblies into isolated particles, dimers, chains, loops and clusters is not merely a convenient image-analysis scheme; it captures the elementary structural motifs that govern the macroscopic mechanical and optical properties of colloidal materials. In particular, during the gelation of short-range attractive colloids, these motifs emerge and evolve dynamically, ultimately forming a spanning network that arrests the system. A fundamental understanding of this process was recently provided by Rouwhorst et al.~\cite{Rouwhorst2020}, who demonstrated that colloidal gelation can be interpreted as a nonequilibrium continuous phase transition, quantitatively described by percolation theory. They showed that the cluster-size distributions follow power laws with exponents –3/2 and –5/2 below and above the percolation threshold, respectively, and that the correlation length diverges with a critical exponent consistent with 3D percolation. This theoretical framework, however, operates at the level of global statistical distributions and does not explicitly resolve the local structural motifs~--- dimers, chains, loops, and branched clusters~--- that constitute the evolving network. Moreover, the competition between bond formation and bond breaking, together with the nonequilibrium nature of the arrest, makes the relative populations of these motifs highly sensitive to the interaction strength, particle shape and assembly kinetics. Therefore, the systematic identification and tracking of these elementary building blocks is not only a prerequisite for validating percolation predictions at the microscopic scale, but also essential for connecting the local connectivity to the global rheological response. This is precisely the gap that the present work addresses.

Beyond the static percolation description, a deeper understanding of gelation requires capturing the kinetic pathways by which clusters grow and evolve. In a parallel study, Rouwhorst et al.~\cite{Rouwhorst2020pre} developed a microscopic kinetic description of this process using a master-equation framework. By analysing the attachment and detachment rates of particles, they demonstrated that, in the weak-attraction regime, single-particle detachment dominates while association rates are essentially independent of cluster size. This allowed them to analytically solve the master kinetic equation and predict the temporal evolution of cluster populations, recovering the power-law mass distributions with exponents -3/2 and -5/2 below and above the percolation threshold. Crucially, their kinetic theory establishes a direct link between the microscopic rates of bond formation and breakage and the macroscopic cluster-size distributions. However, experimental validation and parameterisation of such kinetic models require systematic, statistically meaningful data on the populations of specific structural motifs (dimers, chains, loops, and branched clusters) as a function of time and interaction strength. Manually extracting this information from micrographs is prohibitively time-consuming and subjective. This creates a natural and urgent demand for automated, high-throughput image-analysis tools capable of reliably identifying and counting these elementary building blocks. The present work directly addresses this demand by providing a machine-learning framework for the automatic recognition of colloidal assemblies, thereby enabling quantitative comparisons between experimental observations and the predictions of kinetic cluster-growth theories.

A similar classification was previously considered for spherical particles~\cite{Khusainova2026}, but loops were not identified as a separate group. Loops are important for the mechanical stability of the gel (closed structures increase its elasticity). Here we extend this classification to particles of different shapes~--- ellipsoids, rods, cuboids and spheres~--- and develop a machine‑learning framework for their automated recognition. The central aim is to train a neural network on artificially generated images and then test its generalisation ability on experimental micrographs. Such an approach is particularly valuable for studying rare or difficult‑to‑reproduce assemblies, where large experimental datasets are hardly available.

\section{Methods}
\label{sec:Methods}
\subsection{Neural network model}
An important problem addressed by colloidal analysis is the identification of colloidal assemblies. There are various types of classifications, but here we consider only one of them~\cite{Lotito2017,Lotito2019}. Each assembly is characterized by a certain number of particles and their ``neighbor'', on the basis of which its category is determined. For simplicity, we assume that two particles are neighbors if their surfaces touch. This is a special case, since in a more general formulation one can allow a small gap between two particles, within which they are considered neighbors. Consider the following categories of assemblies~\cite{Lotito2019}:
\begin{figure}[htbp]
\centering
\includegraphics[width=0.8\linewidth]{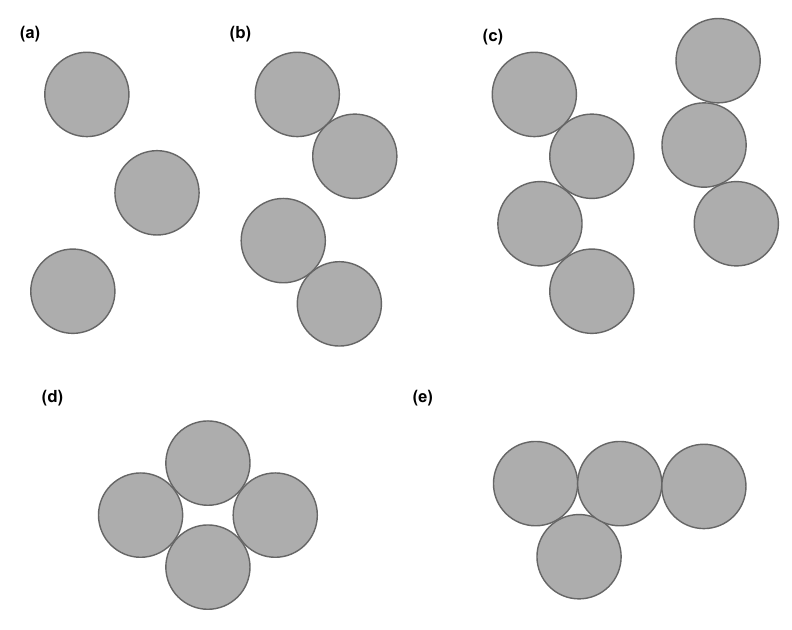}
\caption{Possible colloidal assemblies: a) isolated particles, b) dimers, c) chains, d) loops, and e) clusters (\copyright 2026, Springer Nature~\cite{Khusainova2026})}
\label{fig:Clusters}
\end{figure}
\begin{itemize}[leftmargin=*, label=---]
    \item isolated particles~--- particle $j$ belongs to this group if it has no neighbors (Fig.~\ref{fig:Clusters}a);
    \item dimers~--- two particles $j$ and $k$ belong to this group if $j$ has only neighbor $k$ and $k$ has only neighbor $j$ (Fig.~\ref{fig:Clusters}b);
    \item chains~--- in a chain, all particles have two neighbors, except for the particles at the two ends of the chain, which have only one neighbor (Fig.~\ref{fig:Clusters}c);
    \item loops~--- particles belong to a loop if they all have two neighbors; a loop can be interpreted as a kind of ``closed'' chain or ring (Fig.~\ref{fig:Clusters}d);
    \item clusters~--- particles form a cluster if they belong to an assembly that does not fall into any of the previous categories (Fig.~\ref{fig:Clusters}e).
\end{itemize}

The YOLOv8 model was used as a method to solve the problem. This is a modern model for object detection and classification, developing the ideas of single-stage detectors. It has high speed, a flexible architecture, and high recognition accuracy. For a more detailed description, see our previous work~\cite{Khusainova2026}. The hyperparameters presented in Table~\ref{tab:hyperparameters} were used for training and validation of the models.

\begin{table}[h]
\centering
\caption{Hyperparameters of the model}
\label{tab:hyperparameters}
\begin{tabular}{|l|l|}
\hline
\textbf{Parameter} & \textbf{Value} \\ \hline
Epochs & 100 \\
Image size & 640 \\
Batch size & 3 \\
Optimizer & AdamW \\
Learning rate & 0.001 \\
Momentum & 0.937 \\
Label smoothing & 0.2 \\
Activation function & SiLU \\
Loss & BCE Loss, DFL+CIou Loss \\ \hline
\end{tabular}
\end{table}

\subsection{Dataset}
Datasets play a key role in machine learning, but their generic nature often fails to meet the requirements of highly specialized tasks. In such cases, it becomes necessary to develop specialized datasets tailored to a specific problem. In our previous work~\cite{Khusainova2026}, for spherical particles, we created a dataset primarily based on experimental micrographs (93.5\%) taken from scientific articles. Here, we have created a synthetic dataset because it is problematic to find experimental micrographs in the required quantity for different particle shapes.

In the course of working on the problem, four datasets were artificially created and manually annotated with particles of spherical, ellipsoidal, square, and rod-like shapes. Each of them consists of the corresponding images, ranging from 135 to 155 pieces depending on the specifics of their geometry. To augment the datasets, automatic augmentation was used, where images are rotated by certain angles and various noise is added. The Roboflow service \cite{RoboFlow2025} was used for working with the datasets. All datasets were divided into three sets: training, validation, and test. Each set consists of original images and their corresponding labels. An example of such labels is presented in Table~\ref{tab:labels}.

\begin{table}[h]
\centering
\caption{Structure of data records}
\label{tab:labels}
\begin{tabular}{|c|c|c|c|c|c|}
\hline
Class & $x_1$ & $y_1$ & $x_2$ & $y_2$ & \dots \\ \hline
2 & 0.0971293 & 0.599966 & 0.0971293 & 0.6136681 & \dots \\
1 & 0.7138894 & 0.7144959 & 0.8073184 & 0.7137358 & \dots \\
3 & 0.9428706 & 0.6332427 & 0.9487924 & 0.6381358 & \dots \\
1 & 0.4452560 & 0.6059420 & 0.4393396 & 0.6107149 & \dots \\
2 & 0.7403180 & 0.5285189 & 0.7379487 & 0.5285189 & \dots \\ \hline
\end{tabular}
\end{table}

Each entry in the dataset contains the following structure: first, the polygon class number is indicated (0~--- chain, 1~--- cluster, 2~--- dimer, 3~--- isolated particle, 4~--- loop), followed by the normalized coordinates of the points forming the polygon. The number of points varies depending on the complexity of the annotation of each object. The original images had different sizes, so for standardization, all of them were scaled to 640x640 pixels using the Roboflow platform. Additionally, data augmentation methods were applied to increase the variability of the training set and improve the generalization ability of the model.

In a previous study, it was found that the standard approach using bounding boxes is insufficiently accurate for the analysis of colloidal assemblies of complex shape~\cite{Khusainova2026}. In particular, when individual structures are located close to each other, traditional object identification methods lead to the incorrect merging of independent particles into a single region. This problem is particularly relevant for systems where particles form complex intertwined configurations~--- chains, clusters, or loop-like structures.

To improve the accuracy of object localization in this study, the instance segmentation method with polygonal annotation was applied. This approach makes it possible to precisely determine the boundaries of each object, even in cases of partial overlap or complex morphology of the assemblies. The use of polygons instead of bounding boxes ensures correct separation of closely located particles and accurate correspondence to their real shape, which is critically important for subsequent quantitative analysis of colloidal systems. Additional material, containing the datasets and a script for training models, is freely available~\cite{Khusainova2026github}.

\section{Results and discussion}

\begin{figure}[htbp]
\begin{minipage}[h]{0.49\linewidth}
\center{\includegraphics[width=0.99\linewidth]{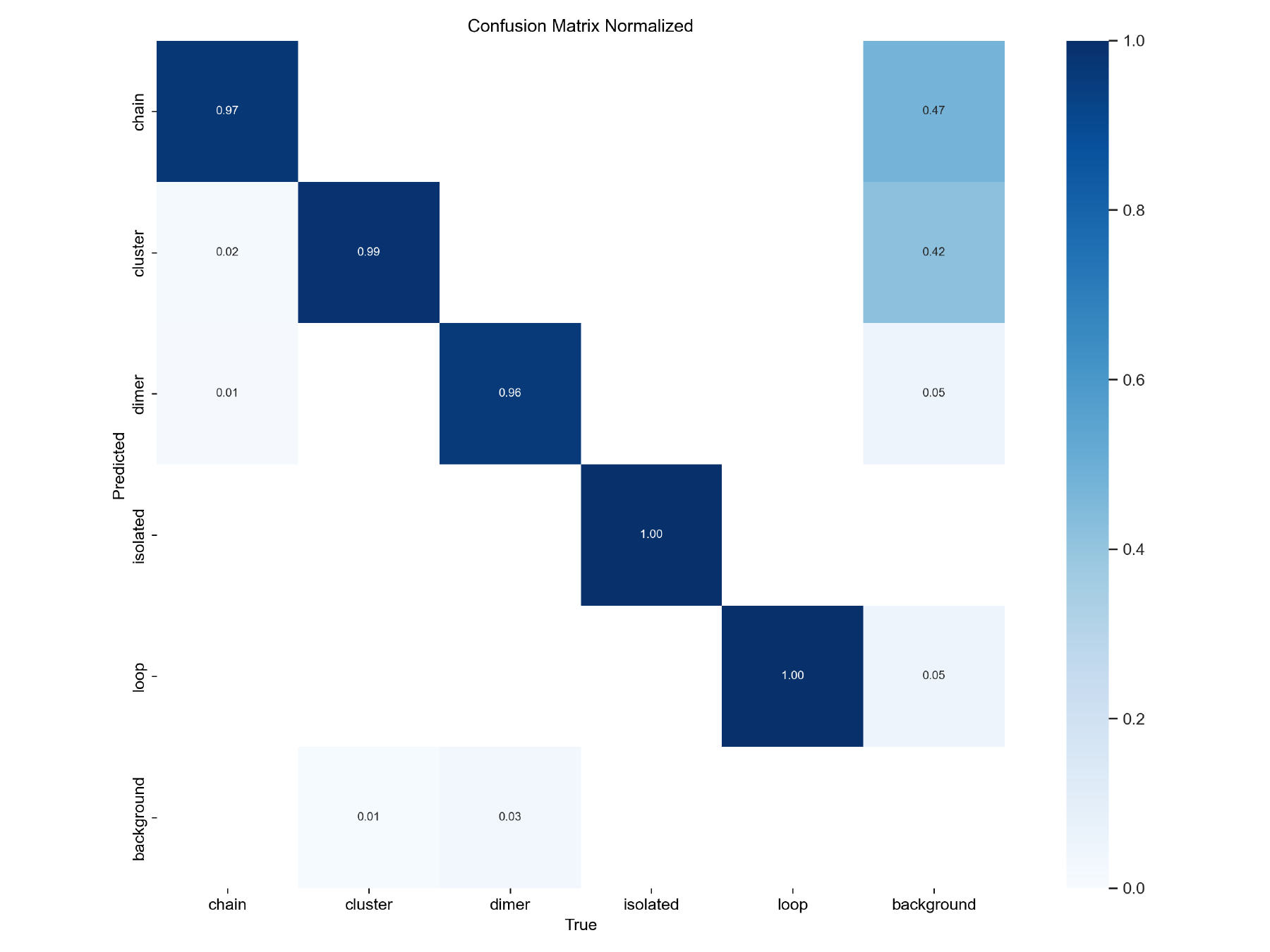} \\ (a)}
\end{minipage}
\hfill
\begin{minipage}[h]{0.49\linewidth}
\center{\includegraphics[width=0.99\linewidth]{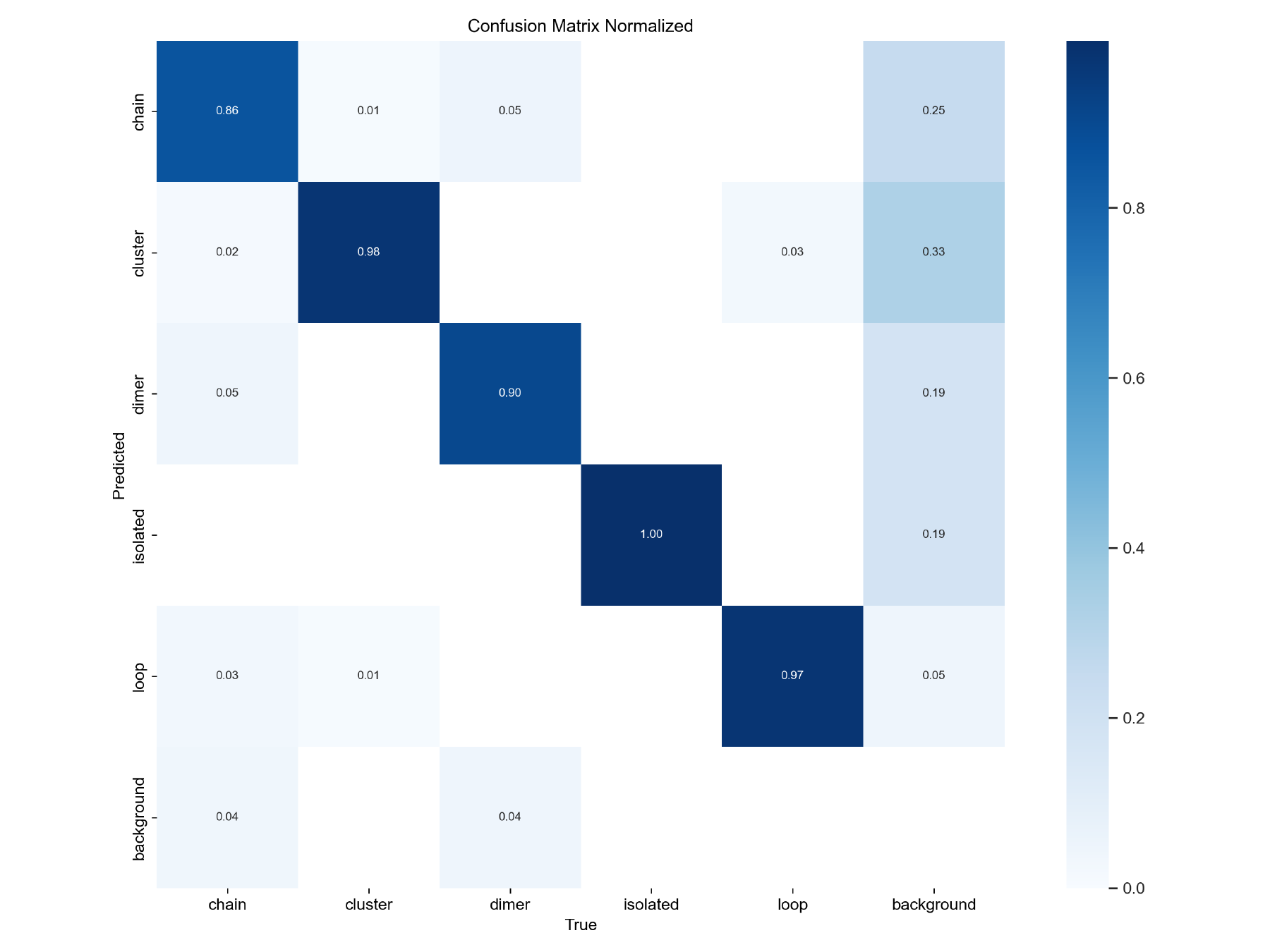} \\ (b)}
\end{minipage}
\\
\begin{minipage}[h]{0.49\linewidth}
\center{\includegraphics[width=0.99\linewidth]{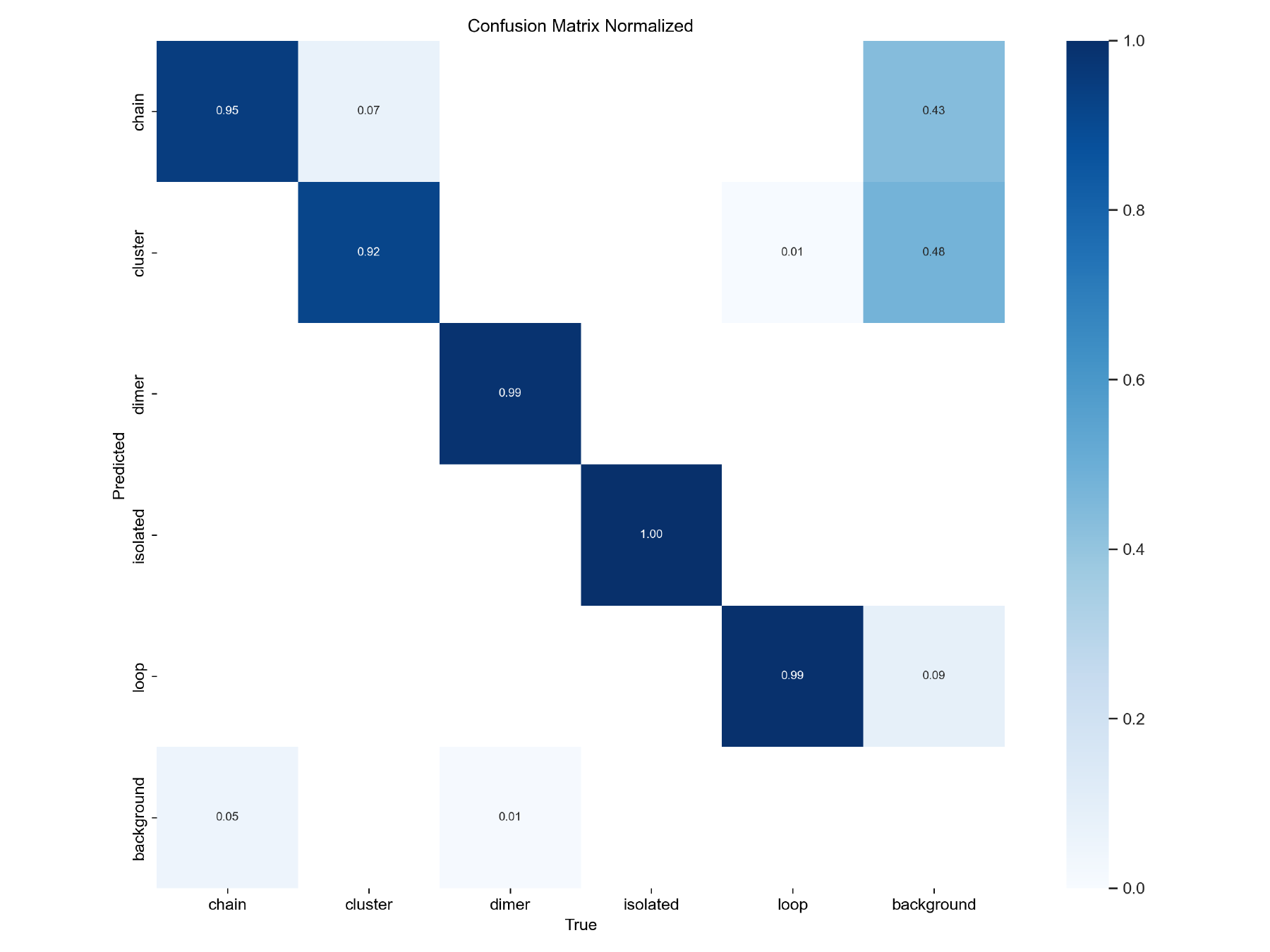} \\ (c)}
\end{minipage}
\hfill
\begin{minipage}[h]{0.49\linewidth}
\center{\includegraphics[width=0.99\linewidth]{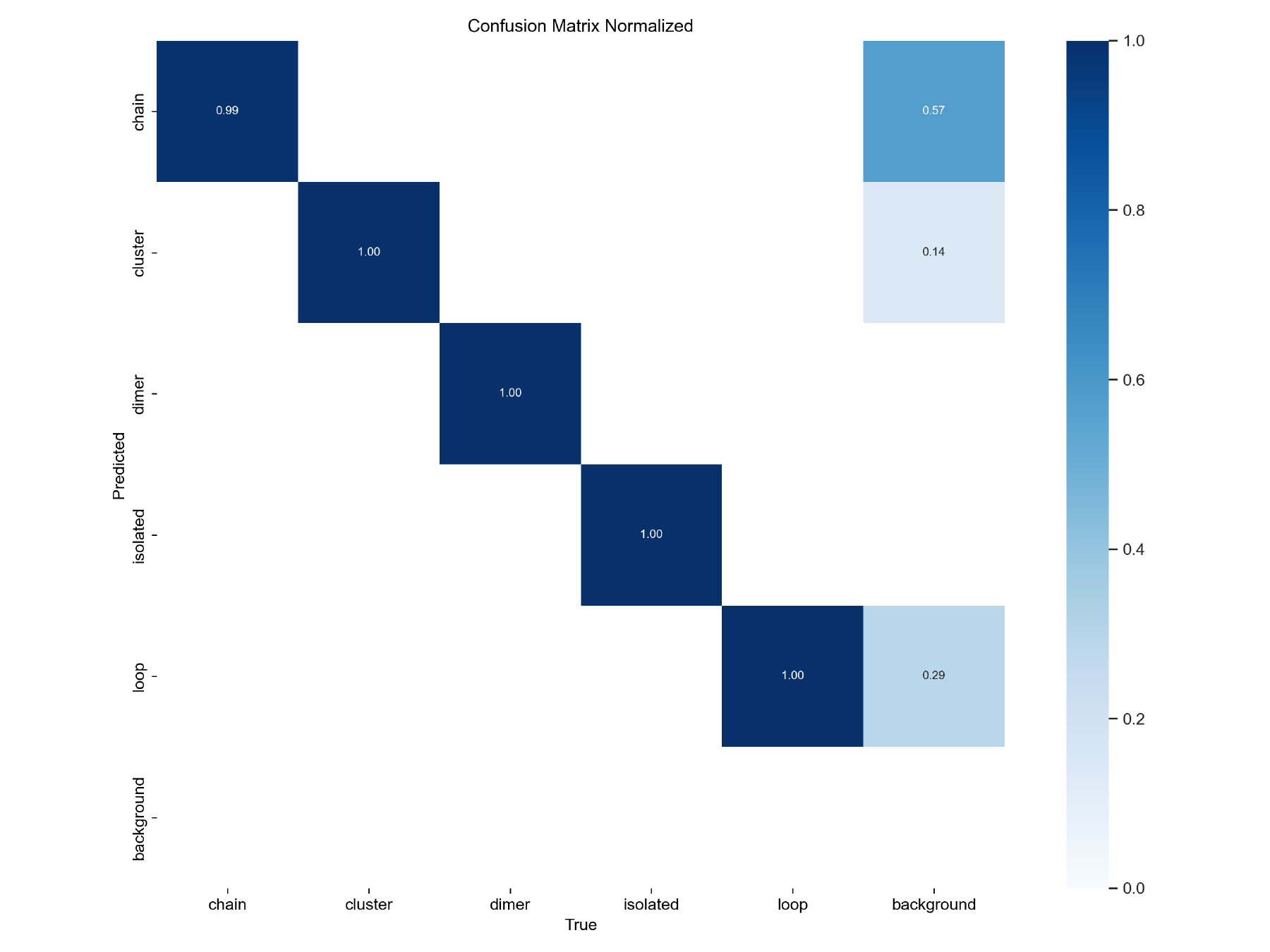} \\ (d)}
\end{minipage}
\caption{Confusion matrices for (a) cuboids, (b) rods, (c) ellipsoids, and (d) spheres}
\label{fig:ConfusionMatrix}
\end{figure}

Model training was performed on a server equipped with an AMD EPYC 7713 processor (64 cores) and eight NVIDIA Quadro RTX A6000 graphics cards. In the course of the study, considering the sizes of the collected datasets, different variants of the YOLOv8-seg architectures were tested: from the lightweight YOLOv8n-seg to the more balanced YOLOv8m-seg. After evaluating the accuracy metrics and training speed, it was found that the YOLOv8m-seg model demonstrates the best balance between performance and segmentation quality. It was precisely the significant expansion of the dataset through augmentation that allowed the YOLOv8m-seg model to show higher accuracy compared to smaller versions of the model. As a result of training, confusion matrices were also constructed (Fig.~\ref{fig:ConfusionMatrix}), clearly showing the distribution of classification errors.

Analysis of the normalized confusion matrices of all four models showed stable classification quality for most classes. However, the background remains the main source of errors, as it is often perceived by the models as certain classes. This may be due to the model mistaking noise or random artifacts in the images for actual assemblies. Despite this, the overall error structure remains similar across all considered cases, indicating the reliability of the models for the main classes and pointing to the need for improved background handling.

\begin{figure}[htbp]
\centering
\includegraphics[width=0.99\linewidth]{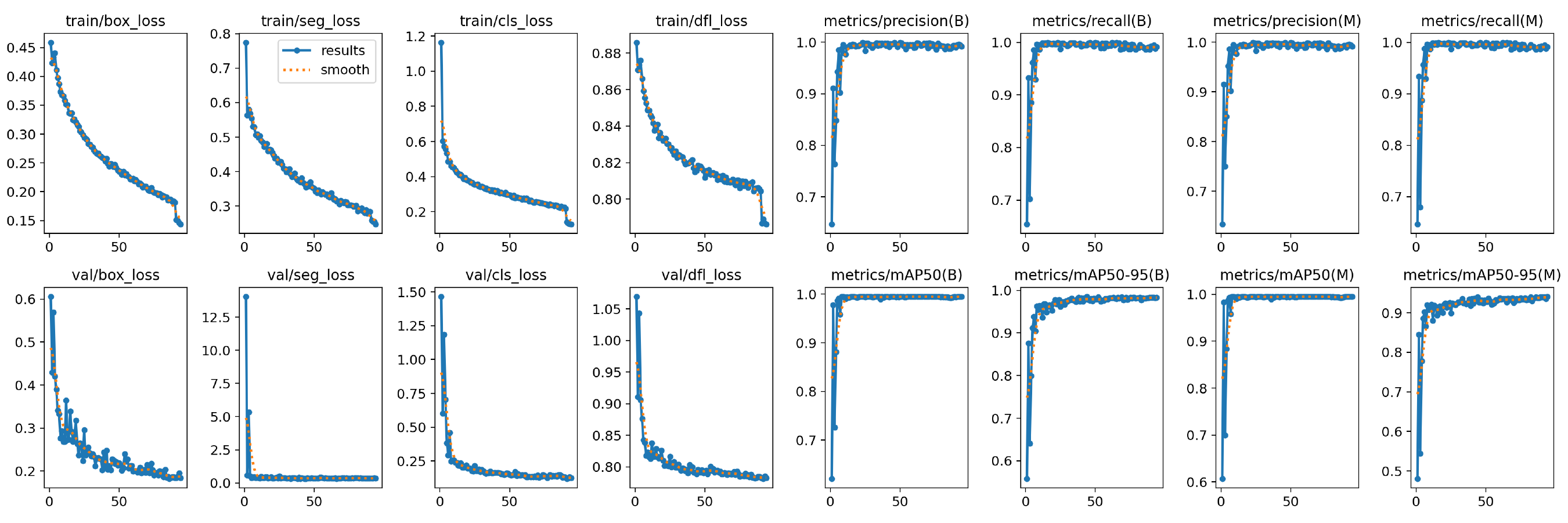}
\caption{Loss curve for spheres}
\label{fig:LossCurves_spheres}
\end{figure}

\begin{figure}[htbp]
\centering
\includegraphics[width=0.99\linewidth]{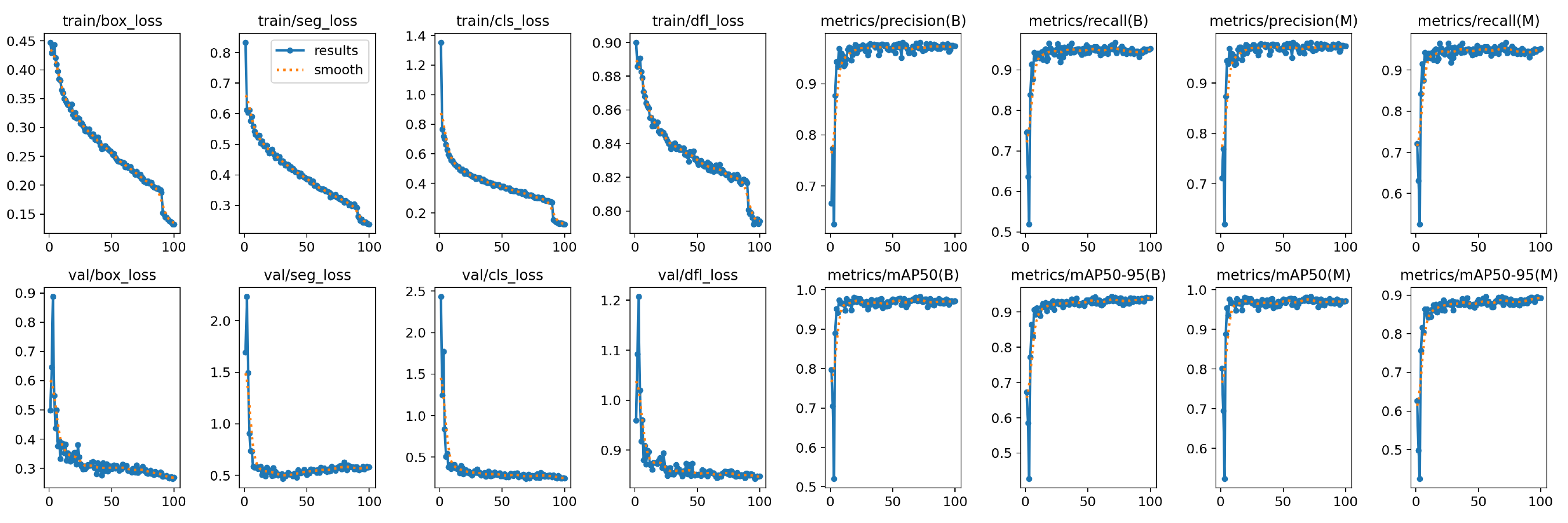}
\caption{Loss curve for ellipsoids}
\label{fig:LossCurves_ellipsoids}
\end{figure}

\begin{figure}[htbp]
\centering
\includegraphics[width=0.99\linewidth]{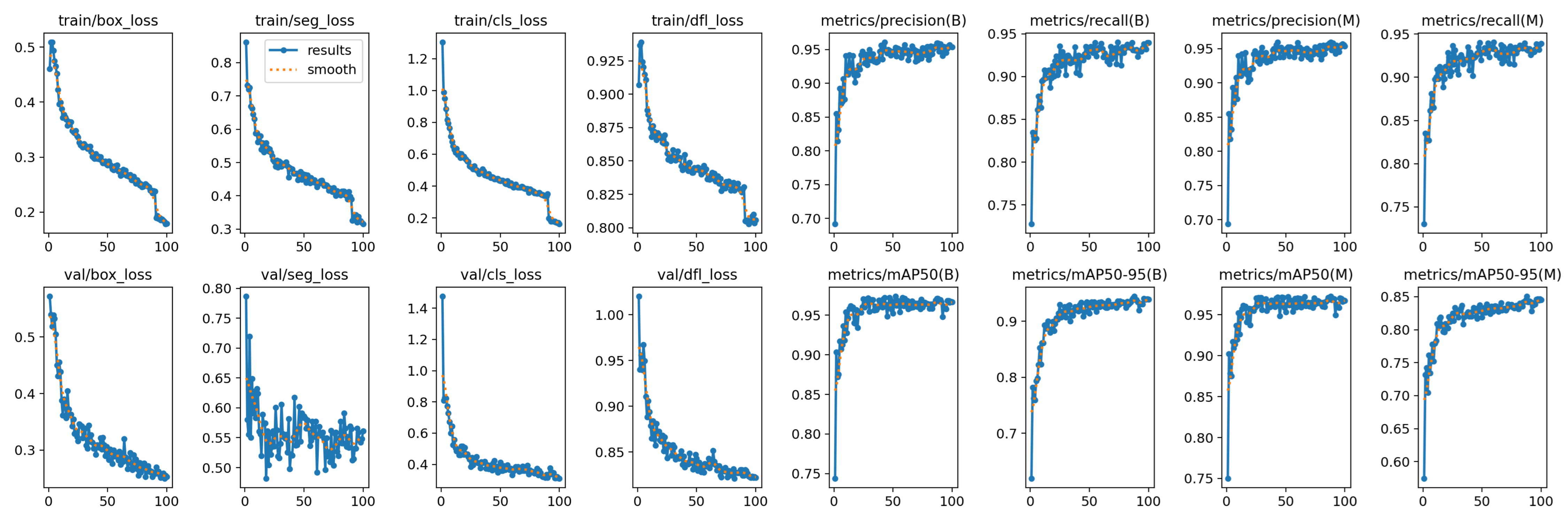}
\caption{Loss curve for rods}
\label{fig:LossCurves_rods}
\end{figure}

\begin{figure}[htbp]
\centering
\includegraphics[width=0.99\linewidth]{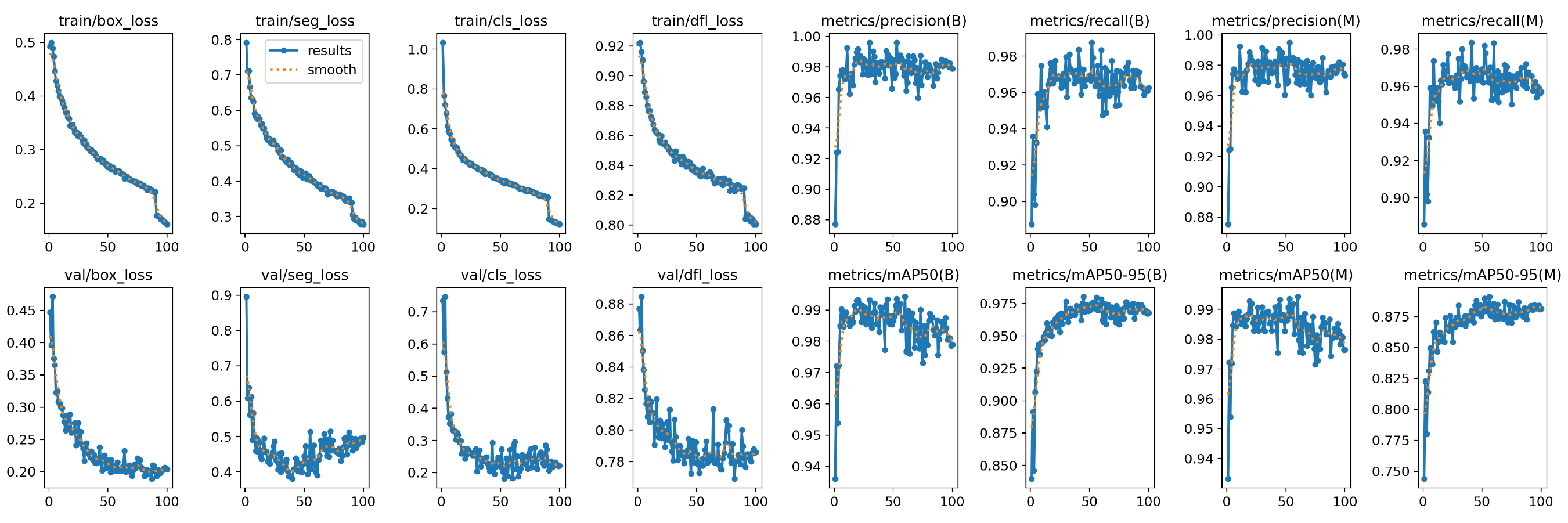}
\caption{Loss curve for cuboids}
\label{fig:LossCurves_cuboids}
\end{figure}

During the study, four specialized models were trained to identify colloidal particles of different shapes: spherical, ellipsoidal, square, and rod-like. Also, to monitor the stability of the training process, loss curves for boxes and masks on the training and validation sets are presented (Fig.~\ref{fig:LossCurves_spheres}--\ref{fig:LossCurves_cuboids}), demonstrating the convergence of the model.

Attention should be paid to the loss curve for cuboids. When analyzing the loss metrics on the validation dataset, signs of overfitting are noticeable: after reaching a minimum at early stages, the value begins to fluctuate and even increases slightly as the number of epochs grows. This indicates that the model begins to overfit to the training data at the expense of generalization ability. In turn, the precision and recall metrics start to slightly decrease towards the end of training, thus demonstrating instability. Therefore, analysis of the plots allows us to conclude that the optimal model quality is achieved approximately at epoch 30--50.

\begin{figure}[htbp]
\centering
\includegraphics[width=0.99\linewidth]{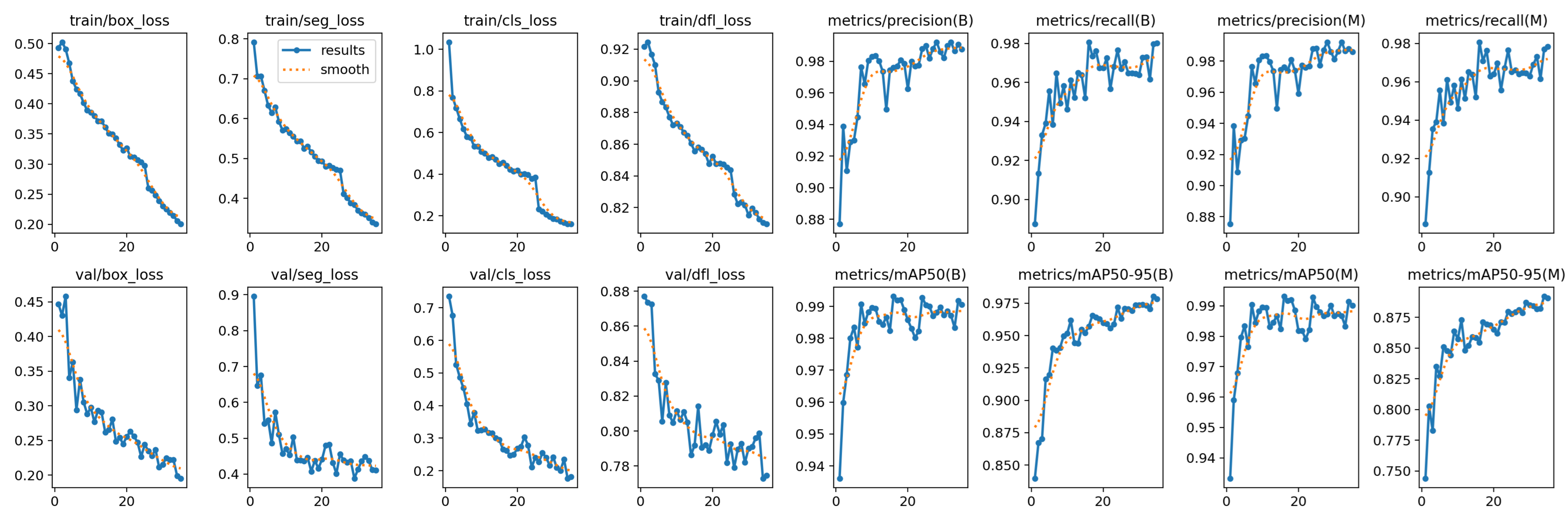}
\caption{Loss curve for cuboids at 35 epochs}
\label{fig:LossCurves_cuboids_35epochs}
\end{figure}

The dataset for cuboids turned out to be the only one of the four in which pronounced signs of overfitting were observed. While on the remaining datasets, the model demonstrated a steady decrease in errors and stable metric values both during training and validation. Due to the identified signs of overfitting, it was decided to limit the training of the model on the cuboid dataset to 35 epochs, which will be sufficient to prevent overfitting (Fig.~\ref{fig:LossCurves_cuboids_35epochs}).

\begin{figure}[htbp]
\centering
\includegraphics[width=0.9\linewidth]{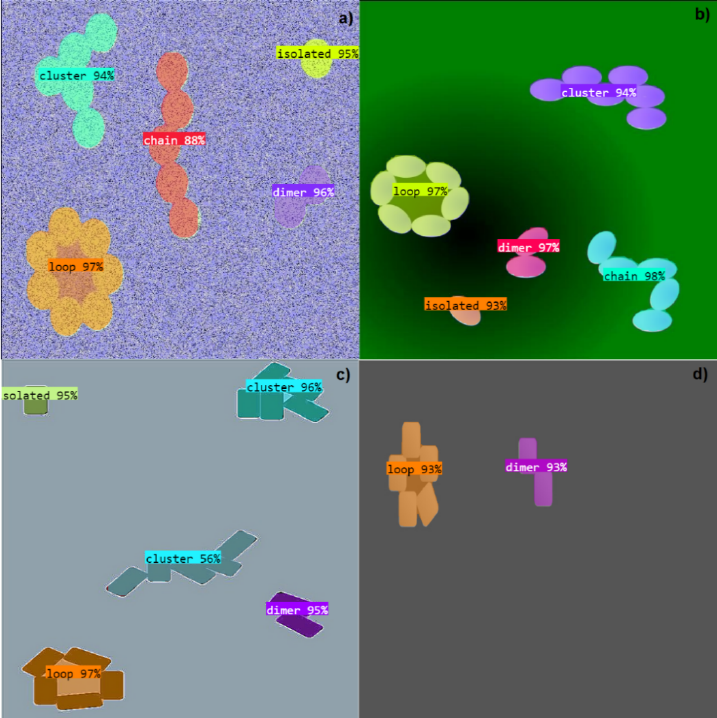}
\caption{Identification results on the test set: a) spherical particles; b) ellipsoidal particles; c) rod-like particles; d) cubic particles.}
\label{fig:IdentificationResults}
\end{figure}

All four models (for different particle shapes) demonstrated approximately the same level of recognition accuracy, 97–99\% (on synthetic data). A preliminary analysis of the models' performance conducted on test sets shows a high percentage of assembly identification, which confirms the stated accuracy (Fig.~\ref{fig:IdentificationResults}).

However, to achieve the stated goal, it is necessary to analyze the trained models on experimental micrographs. The results of the analysis of the four trained models on a sample of forty experimental images (ten images for each particle shape) revealed the following key identification troubles:

\begin{enumerate}[leftmargin=*]
    \item excessive identification of artifacts (the model may sometimes identify extra objects such as debris, labels, background noise, etc., as particle assemblies);
    \item missing target assemblies (some significant assemblies remain undetected);
    \item misclassification of assemblies (the model incorrectly determines the types of assemblies);
    \item probability overlap (conflicts in areas of overlapping predictions);
    \item particle truncation (loss of particle fragments, i.e., parts of their area, in assemblies).
\end{enumerate}

\begin{figure}[htbp]
	\centering
	\begin{minipage}{0.48\linewidth}
		\centering
		\includegraphics[width=\linewidth]{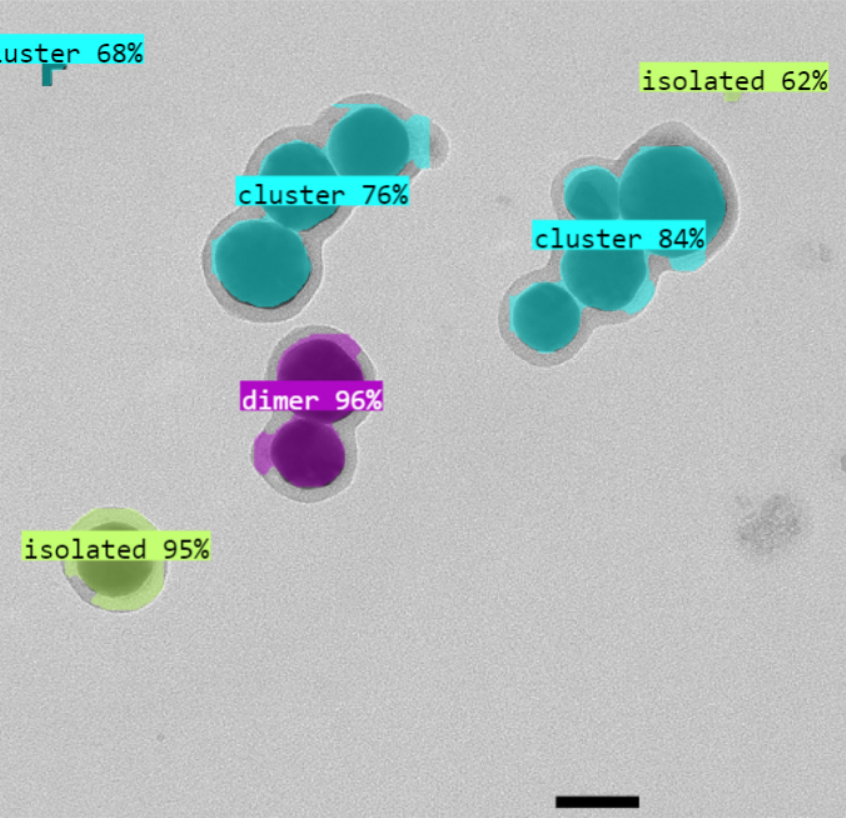}\\(a)
	\end{minipage}
	\hfill
	\begin{minipage}{0.48\linewidth}
		\centering
		\includegraphics[width=\linewidth]{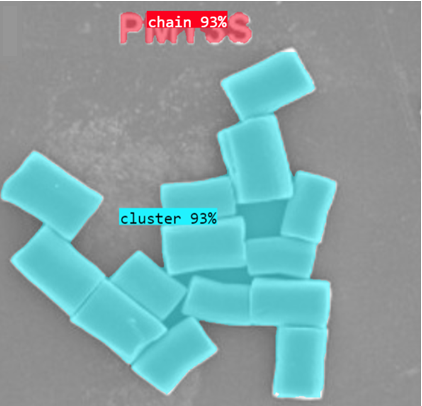}\\(b)
	\end{minipage}
	\caption{Over-identification: (a) spheres (\textcopyright~Busch et al.~\cite{Busch_2019}, ACS AuthorChoice) and (b) rods (\textcopyright~2023 Wiley-VCH GmbH~\cite{Li2023})}
	\label{fig:OverIdentification}
\end{figure}

The observed over-identification of clusters (Fig.~\ref{fig:OverIdentification}) can be explained by the limited variability of the synthetic training dataset. Since the dataset was created from artificial data, it does not contain noise, artifacts, texture features, or random labels and marks that are typical for real images. This has led to the model lacking robustness to such distortions, which manifests as false positives on real micrographs.

\begin{figure}[htbp]
	\centering
	\begin{minipage}{0.48\linewidth}
		\centering
		\includegraphics[width=\linewidth]{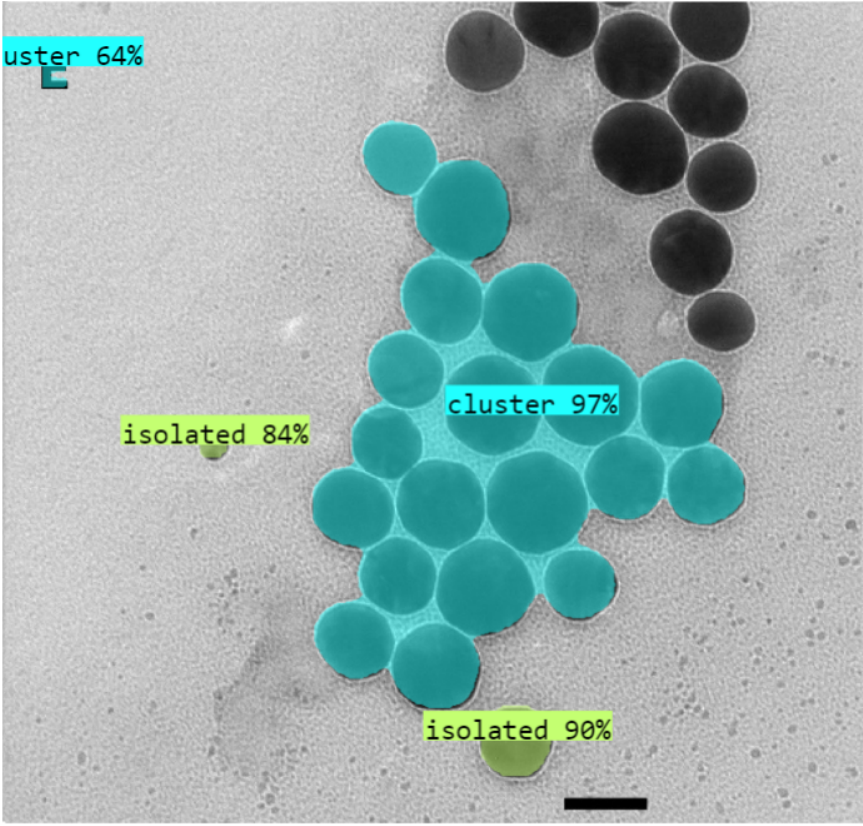}\\(a)
	\end{minipage}
	\hfill
	\begin{minipage}{0.48\linewidth}
		\centering
		\includegraphics[width=\linewidth]{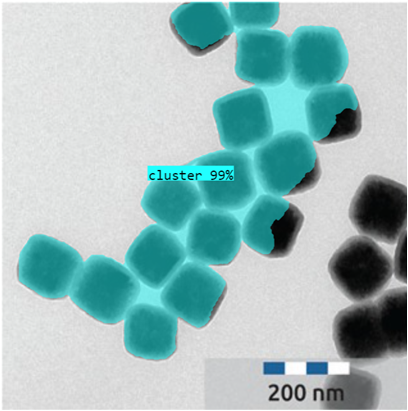}\\(b)
	\end{minipage}
	\caption{Omission of target assemblies: (a) spheres (\textcopyright~Busch et al.~\cite{Busch_2019}, ACS AuthorChoice) and (b) cuboids (\textcopyright~Rosenberg et al. 2020~\cite{Rosenberg2020}, CC BY-NC 3.0)}
	\label{fig:MissedObjects}
\end{figure}

    The omission of target assemblies (Fig.~\ref{fig:MissedObjects}) may be related to an insufficient number of examples of complex cases in the training dataset. It is possible that the data on which the model was trained lacks or has an insufficient representation of complex and non-obvious object examples, which leads to problems with the model's generalization ability.

\begin{figure}[htbp]
    \centering
    \includegraphics[width=0.6\linewidth]{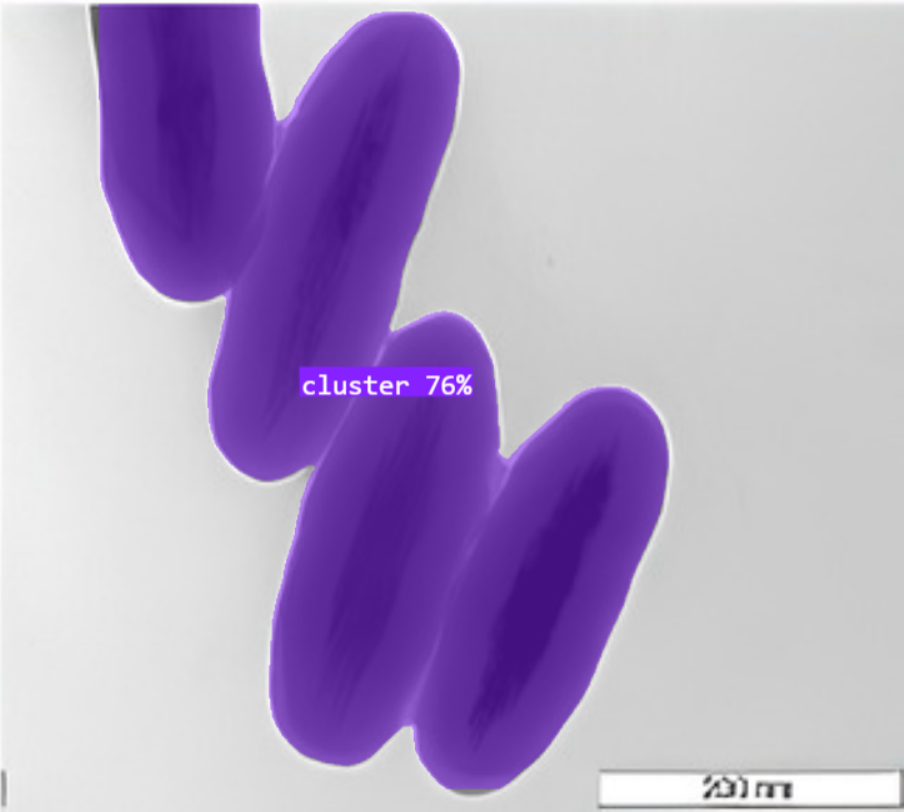}
    \caption{Misclassification of an assembly (\copyright 2006 American Chemical Society)~\cite{Sacanna_2006}. }
    \label{fig:ErroneousClassification}
    \end{figure}

    The problem of misclassification of assemblies (Fig.~\ref{fig:ErroneousClassification}) may be caused by insufficient feature distinguishability: similar morphological characteristics among different types of assemblies. For example, one can notice that a cluster and a chain are quite similar. The difference may literally be in a single particle: the addition of just one neighbor turns a chain into a cluster. It is likely that the training data did not contain a sufficient number of such ``borderline'' examples where it is difficult to unambiguously determine the type of assembly. Proper recognition of these motives is critical for verifying the predictions of percolation theory. Such errors (for example, confusion between chains and clusters) can lead to distortions in power-law distributions~\cite{Rouwhorst2020}.

\begin{figure}[htbp]
    \centering
    \includegraphics[width=0.6\linewidth]{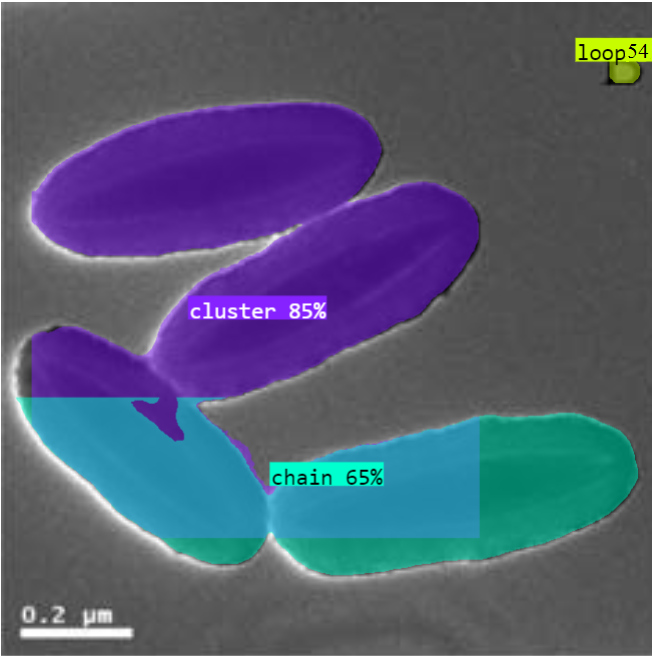}
    \caption{Probability overlap (\copyright 2010 The Royal Society of Chemistry)~\cite{Li_2010}.}
    \label{fig:ProbabilityOverlap}
\end{figure}

The problem of probability overlap (Fig.~\ref{fig:ProbabilityOverlap}) may be related to feature ambiguity. When the shapes of objects are too similar, the model may have difficulty making an unambiguous decision about which class to assign the object to. This is due to the fact that the features may be so similar that the algorithm cannot confidently draw a distinction between different types.

\begin{figure}[htbp]
    \centering
    \includegraphics[width=0.6\linewidth]{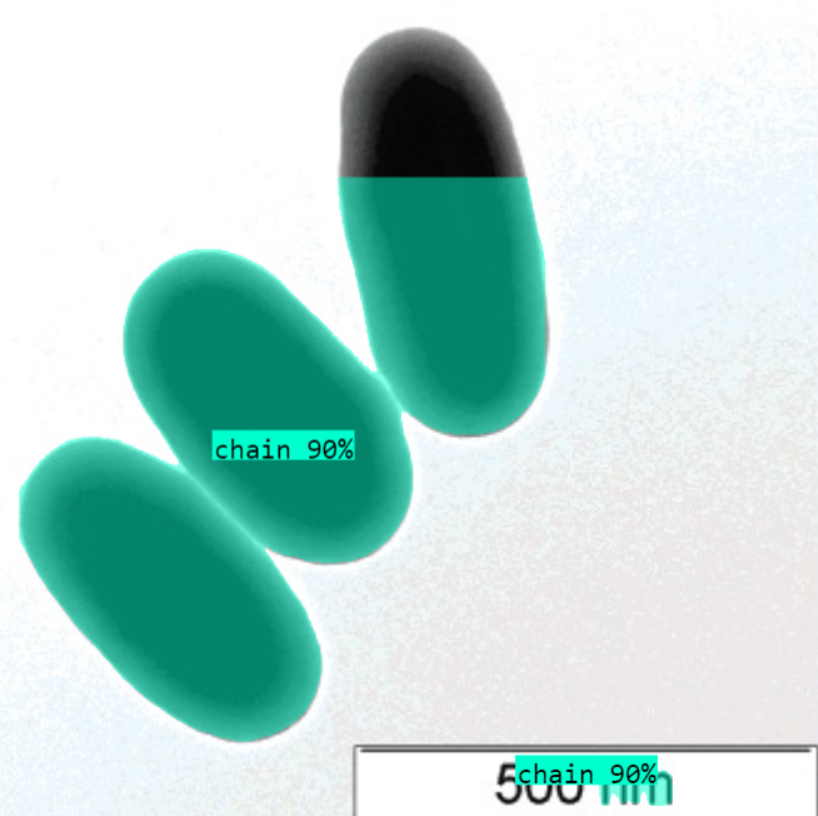}
    \caption{Particle truncation (\copyright 2006 American Chemical Society)~\cite{Sacanna_2006}.}
    \label{fig:Cutting}
\end{figure}

The problem of truncation of particle fragments (Fig.~\ref{fig:Cutting}) in assemblies is not completely understood. This effect requires attention, as it can distort the results of qualitative object evaluation. Furthermore, in Fig.~\ref{fig:Cutting}, the label ``500 nm'' is recognized as a chain (erroneous classification).

Given the multifaceted nature of the identified problems, using a single metric to evaluate error is methodologically incorrect. Different types of errors require specialized approaches. In all cases, except for the situation with probability overlap, the application of standard formulas for calculating precision and recall is acceptable:

\begin{equation}
\text{Precision} = \frac{\text{TP}}{\text{TP} + \text{FP}},
\label{eq:precision}
\end{equation}
where TP~--- true positive objects (correctly assigned object to a class), FP~--- false positive objects (incorrectly assigned object to a class);

\begin{equation}
    \text{Recall} = \frac{\text{TP}}{\text{TP} + \text{FN}},
\label{eq:recall}
\end{equation}
where FN~--- false negative objects (it is incorrectly asserted that the object does not belong to the class).

However, for the case of probability overlap, these formulas are not applicable, since they are not intended for soft classification tasks. Given the detected identification problems, these metrics should be calculated not only for assemblies but also for all particles in the images. This will ensure the completeness of the evaluations.

Soft classification is an approach in machine learning where, instead of clearly assigning an object to a class, the model returns probabilities of the object belonging to different classes. This directly relates to cases of probability overlap. To calculate precision and recall in this case, the following formulas are applied \cite{Fr_nti_2023}:

\begin{equation}
    \text{Soft Precision} = \frac{\text{card}(G \cap P_i)}{\text{card}(P_i)} ,
    \label{eq:soft_precision}
\end{equation}

\begin{equation}
    \text{Soft Recall} = \frac{\text{card}(G \cap P_i)}{\text{card}(G)},
    \label{eq:soft_recall}
\end{equation}
where $G$ is the ground truth, truly positive or annotated data, $P_i$ is the predicted items, $\text{card}(G \cap P_i)$ is the sum of all probabilities of true model predictions, $\text{card}(P_i)$ is the sum of all probabilities of model predictions (both true and false), $\text{card}(G)$ is the number of true assemblies in the image (the probability of each true assembly equals 1). Here, the operator $\text{card}$ should be understood as soft cardinality~\cite{Fr_nti_2023}.

Thus, for calculating precision and recall, not only entire assemblies but also individual particles in the images are taken into account. During the analysis of forty experimental images (ten pictures for each trained model), the evaluation metrics presented in Table~\ref{tab:comparisonOnRealData} were obtained. In this table, the indices $c$ and $p$ denote the relation to configuration and to particle, respectively. In cases where probability overlap occurred, formulas \eqref{eq:soft_precision} and \eqref{eq:soft_recall} were used; otherwise, formulas \eqref{eq:precision} and \eqref{eq:recall} were used. The average values of Precision and Recall were determined across all ten images.

For comparison with experimental data, a similar procedure was performed on simulation images taken from the test set. The processing of forty synthetic images (ten pictures for each trained model) was carried out according to the same principle. The choice of formulas \eqref{eq:precision}, \eqref{eq:recall} or \eqref{eq:soft_precision}, \eqref{eq:soft_recall} depended on the presence of probability overlap. The averaged Precision and Recall values presented in Table~\ref{tab:comparisonOnSyntheticData} allowed a direct comparison with the metrics obtained on real data (Table~\ref{tab:comparisonOnRealData}). Details of our calculations and additional statistical data are available in the accompanying materials (see Supplementary Information).

\begin{table}[h!]
\centering
\caption{Comparative analysis of the application of models to experimental data}
\label{tab:comparisonOnRealData}
\begin{tabular}{|l|c|c|c|c|}
\hline
Particle shape & Precision\textsubscript{c} & Recall\textsubscript{c} & Precision\textsubscript{p} & Recall\textsubscript{p} \\
\hline
Spheres       & 0.73 & 0.88 & 0.75 & 0.84 \\
Cuboids     & 0.38 & 0.38 & 0.44 & 0.44 \\
Ellipsoids  & 0.45 & 0.61 & 0.48 & 0.61 \\
Rods     & 0.31 & 0.65 & 0.38 & 0.65 \\
\hline
\end{tabular}
\end{table}

\begin{table}[h!]
\centering
\caption{Comparative analysis of the application of models to synthetic data}
\label{tab:comparisonOnSyntheticData}
\begin{tabular}{|l|c|c|c|c|}
\hline
Particle shape & Precision\textsubscript{c} & Recall\textsubscript{c} & Precision\textsubscript{p} & Recall\textsubscript{p} \\
\hline
Spheres       & 1 & 1 & 1 & 1 \\
Cuboids     & 0.98 & 1 & 0.97 & 1 \\
Ellipsoids  & 0.96 & 0.99 & 0.94 & 0.99 \\
Rods     & 0.94 & 1 & 0.94 & 1 \\
\hline
\end{tabular}
\end{table}

A comparison of the obtained metrics between experimental and synthetic data demonstrates a significant difference in model performance. On synthetic data, the model shows nearly ideal results. For spheres, all Precision and Recall metric values are unity. For ellipsoids, cuboids, and rods, the indicators decrease slightly, remaining in the range of 0.94--1. At the same time, on experimental data, significantly lower performance is observed. Even for well-recognized spheres, the values of $\text{Precision}_{c}$ and $\text{Recall}_{c}$ are only 0.73 and 0.88, respectively. In the case of complex shapes such as rods, the identification accuracy drops to 0.31. This discrepancy in results indicates a substantial difference between idealized synthetic data and real experimental conditions, where various noises, artifacts, and complexities associated with the specific shapes of particles are present.

\begin{table}[h!]
	\centering
	\begin{threeparttable}
		\caption{The relative error of the indicators on experimental data with respect to the synthetic benchmark (\%)}
		\label{tab:relativeErrorAnalysis}
		\begin{tabular}{|l|c|c|c|c|c|}
			\hline
			Particle shape & $\delta(P_c)$ &  $\delta(R_c)$ &  $\delta(P_p)$ &  $\delta(R_p)$ & \textbf{Average, \%}\tnote{*} \\ \hline
			Spheres        & 27.0 & 12.0 & 25.0 & 16.0 & \textbf{20.0} \\
			Cuboids      & 61.2 & 62.0 & 54.6 & 56.0 & \textbf{58.5} \\
			Ellipsoids   & 53.1 & 38.4 & 48.9 & 38.4 & \textbf{44.7} \\
			Rods      & 67.0 & 35.0 & 59.6 & 35.0 & \textbf{49.2} \\ \hline
			\multicolumn{5}{|l|}{\textbf{Average for all forms}} & \textbf{43.1} \\ \hline
		\end{tabular}
		\begin{tablenotes}
			\small
			\item[*] It is calculated as the arithmetic mean of the relative errors of four metrics.
		\end{tablenotes}
	\end{threeparttable}
\end{table}

A comparative analysis of the error $\delta$ (Table~\ref{tab:relativeErrorAnalysis}) shows a significant degradation of metrics when moving from synthetic data to experimental data, $$\delta(X)=\frac{\left| X_e - X_s \right|}{X_s}\times 100\%,$$ where $X_e$ and $X_s$ are the values of a certain metric obtained on experimental and synthetic images. In Table~\ref{tab:relativeErrorAnalysis}, the notations $P$ and $R$ are used for Precision and Recall, respectively, for brevity. The average relative deviation $\delta$ across all object types was 43.1\%. The ``Spheres'' class demonstrates the best robustness (average deviation of 20\%), which is explained by the geometric simplicity of the shape and lower sensitivity of the algorithm to the imaging angle. The greatest difficulties are caused by ``Cuboids'' (error of 58.5\%) and ``Rods'' (49.2\%). Such a wide spread of values (from 20\% to 58.5\%) indicates that the synthetic sample does not fully reproduce the optical artifacts and shadowing features characteristic of angular and elongated objects in real conditions. The high error value in Recall metrics for cuboids (62.0\%) relative to synthetic data indicates missed detections of objects with complex geometry when processing experimental frames.

A detailed analysis of the metrics revealed an imbalance between precision and recall. The average relative error of Precision metrics (52\% for configurations and 47\% for particles) significantly exceeds the error of Recall (36.85\% for configurations and 36.35\% for particles). This indicates that the main problem of the model when transitioning to real data is the increase in the number of false positives (erroneous detection of background noise or artifacts as particles and their assemblies). At the same time, the model demonstrates greater robustness in terms of missed objects: the algorithm finds particles and their assemblies in a real environment relatively reliably, although not as well as on synthetic data. However, it makes mistakes much more often in classifying their configurations. A particularly critical gap is observed for the ``Rods'' class, where when determining configurations, precision drops by 67\%, while recall drops by only 35\%. In the case of cuboids, the degradation of precision and recall occurs approximately equally when transitioning to real data.

To rationalize the observed differences in identification accuracy across particle shapes (Table~\ref{tab:shape_characteristics}), we correlate the error rates with key morphological descriptors, such as the aspect ratio (the ratio of the particle length to its width) and shape complexity. While spherical particles (aspect ratio = 1) exhibit the lowest average error of 20\%, the error increases for elongated shapes like rods (aspect ratio $>$ 3, error 49.2\%). This trend is consistent with the known sensitivity of colloidal assembly and image analysis to particle anisotropy. However, the highest error is observed for cuboids (58.5\%), despite their moderate aspect ratio. This discrepancy can be attributed to the presence of sharp edges and flat facets, which introduce significant optical artifacts (e.g., shadows and highlights) in experimental micrographs that are not replicated in our synthetic training dataset. This finding highlights that not only the aspect ratio, but also the overall shape complexity and the resulting imaging signatures, critically affect the performance of machine-learning models, a factor that has been previously emphasized in the context of pattern detection in colloidal assemblies~\cite{Moucka2005,Reis2006,Dillmann2008,Lotito2016,Lotito2018,Lotito2020}.

\begin{table}[htbp]
	\centering
	\caption{Geometric characteristics of particles and their relation to recognition complexity}
	\label{tab:shape_characteristics}
	\begin{tabularx}{\columnwidth}{|>{\hsize=0.75\hsize}X|>{\hsize=0.65\hsize}X|>{\hsize=1.85\hsize}X|>{\hsize=0.75\hsize}X|}
		\hline
		\textbf{Particle shape} & \textbf{Aspect ratio} & \textbf{``Angularity'' / Shape complexity} & \textbf{Average error, \%} \\
		\hline
		Spheres                 & 1.0                            & Low (isotropic, no facets) & 20.0 \\
		Cuboids                 & 1.0--2.0                      & Very high (flat facets and sharp edges cause shadows and highlights) & 58.5 \\
		Ellipsoids              & 1.5--2.0                      & Moderate (smooth anisotropy) & 44.7 \\
		Rods                    & $>$ 3.0                        & High (strong elongation, orientation-dependent) & 49.2 \\
		\hline
	\end{tabularx}
\end{table}

The created datasets and trained models are freely available for use~\cite{Khusainova2026github}. The corresponding modules have been integrated into the previously developed information system \cite{Khusainova2025} (\href{https://isanm.space/}{https://isanm.space/}).

\section*{Conclusion}

During this work, colloidal assemblies of particles of various shapes (spheres, cuboids, rods, and ellipsoids) were investigated with the aim of training a neural network model for their automatic identification. To solve this problem, the YOLOv8 model was selected and analyzed. To train the models, datasets containing synthetic images of assemblies were created and annotated.

The YOLOv8 model demonstrated high recognition accuracy on synthetic images, but its performance on experimental data decreased significantly (by 43.1\% on average). This is likely due to the limitations of the synthetic dataset, which does not fully reproduce the specifics of real experimental micrographs, as well as the imperfection of the shapes of real objects. The shape of real particles often differs significantly from the ideal geometric models used in the datasets. For example, particles labeled as ellipsoidal may in practice have an elongated shape resembling rods. Some particles have ambiguous shapes~--- they are difficult to accurately classify as, for instance, rods or cuboids, especially in borderline cases. After all, a cuboid is a special case of a rod, and a sphere is a special case of an ellipse. Rod-like particles are sometimes curved.

The obtained results indicate that datasets based exclusively on artificially generated data are insufficiently generalizable for working with real images. 
We also found a clear correlation between the particle shape complexity and the model performance: the identification error increases from 20\% for isotropic spheres to 49--58\% for anisotropic and faceted particles (rods and cuboids). This trend underscores the importance of accounting for geometrical descriptors (such as aspect ratio and angularity) when designing training datasets for machine-learning-based analysis of colloidal assemblies.
To improve recognition accuracy, it is necessary to create a new training dataset using experimental micrographs. In the future, the results of new models can be compared with the results of our models (see \href{https://isanm.space/}{https://isanm.space/}) using specific images as examples. Also of interest are mixtures of particles of different shapes. The datasets developed here and the trained models are freely available~\cite{Khusainova2026github} and can be useful to engineers and researchers working in related fields.

The substantial gap in model performance between synthetic and experimental data underscores the challenges in transferring machine-learning models trained on idealised images to real experimental conditions. Nevertheless, the framework established here represents a critical first step towards the automated extraction of structural information required by modern kinetic theories of gelation~\cite{Rouwhorst2020pre}. With future improvements based on experimentally annotated datasets, our approach has the potential to provide the high-throughput, quantitative data on motif populations needed to validate and refine master-equation descriptions of cluster growth, ultimately bridging the gap between microscopic dynamics and macroscopic rheological response.

\section*{Acknowledgement}
This work is supported by Grant No. 22-79-10216 from the Russian Science Foundation (\href{https://rscf.ru/en/project/22-79-10216/}{https://rscf.ru/en/project/22-79-10216/}).

  \bibliographystyle{elsarticle-num}
  \bibliography{Khusainova_et_al2025}





\end{document}